\documentclass[a4paper,11pt]{article}

\usepackage{jcappub} 

\usepackage{graphicx}
\usepackage{amssymb}
\usepackage{ amssymb }

\input epsf.sty

\newcommand{\bd}{\begin{displaymath}}
\newcommand{\ed}{\end{displaymath}}
\newcommand{\ba}{\begin{eqnarray}}
\newcommand{\ea}{\end{eqnarray}}
\newcommand{\be}{\begin{equation}}
\newcommand{\ee}{\end{equation}}
\newcommand{\ben}{\begin{eqnarray}}
\newcommand{\een}{\end{eqnarray}}

\title{\boldmath The 21-cm Signal from the Cosmological Epoch of Recombination}
\author[a,c]{A. Fialkov}
\author[b,c]{and A. Loeb}

\affiliation[a]{Departement de Physique, Ecole Normale Superieure, CNRS,\\ 24 rue Lhomond, 75005 Paris, France}
\affiliation[b]{Department of Astronomy, Harvard University,\\60 Garden 
Street, MS-51, Cambridge, MA, 02138, USA}
\affiliation[c]{The Raymond and Beverly Sackler School of Physics and Astronomy, Tel Aviv University,\\Tel Aviv 69978, Israel}

\emailAdd{anastasia.fialkov@phys.ens.fr; aloeb@cfa.harvard.edu}

\abstract{The redshifted 21-cm emission by neutral hydrogen offers a
unique tool for mapping structure formation in the early universe in
three dimensions.  Here we provide the first detailed calculation of
the 21-cm emission signal during and after the epoch of hydrogen
recombination in the redshift range of $z\sim 500$--$1,100$,
corresponding to observed wavelengths of 100--230 meters. The 21-cm
line deviates from thermal equilibrium with the cosmic microwave
background (CMB) due to the excess Ly$\alpha$ radiation from hydrogen
and helium recombinations.  The resulting 21-cm signal reaches a brightness temperature
of a milli-Kelvin, orders of magnitude larger than previously estimated. Its
detection by a future lunar or space-based observatory could improve
dramatically the statistical constraints on the cosmological initial
conditions compared to existing two-dimensional maps of 
the CMB anisotropies.  }

\begin{document}
\maketitle
\flushbottom

\section{Introduction}

Hydrogen atoms provide an excellent tracer for the evolution of
structure from the epoch of recombination at $z\sim 1,100$, when these
atoms first formed, to the epoch of reionization  at $z\sim 7$, when they were
broken to their constituent electrons and protons \cite{Loeb:2012}. 
The 21-cm transition between the two hyperfine levels of the 1S ground
state has a small optical depth, and so the signal from each redshift
$z$ can be mapped independently by measuring the extra-Galactic
emission at redshifted wavelengths of $21\times(1+z)$ cm \cite{Pritchrd:2011, Furlanetto:2006}. 
Since the 21-cm signal is coupled to astrophysical sources and
cosmological parameters at the emission redshift, it is expected to
provide precious information about the conditions in the early
Universe.

A major effort is currently underway to detect the 21-cm signal from
the epoch of reionization in the redshift range of $6 < z < 15$
corresponding to observed wavelengths of $\sim 1.5$--$7$ meters. The
newly constructed arrays of radio dipole antennae include the
Murchison Widefield Array (MWA) \cite{MWAref}, the LOw Frequency ARray
(LOFAR) \cite{LOFARref}, the Precision Array for Probing the Epoch of
Reionization (PAPER) \cite{PAPER},
and the Large Aperture Experiment to Detect the Dark Ages (LEDA)
\cite{LEDA:2012}, and plans exist for the next generation experiments
that may reach up to $z\sim 30$, such as the Hydrogen Epoch of
Reionization Array (HERA)\footnote{http://reionization.org/}
the Square Kilometer Array (SKA) \cite{Carilli:2004} and the Dark
Ages Radio Explorer (DARE) \cite{Burns:2012}.

The primary challenge for detecting the redshifted 21-cm signal
involves the foregrounds of Galactic and extragalactic synchrotron
emission which brightens with decreasing photon frequency.  The
foreground emission at very low frequencies is poorly constrained, and
in fact, there are no available maps of all-sky emission at
frequencies below 30 MHz. The state of the art in this field are maps
of a part of the southern sky at resolution of 5$^\circ$--30$^\circ$
measured from the ground \cite{RadioMaps1} and maps produced by the
Radio Astronomy Explorer-2 satellite with the resolution of
$>30^\circ$ made in the 1970s \cite{RadioMaps2}. Only few space-based
observations were conducted so far in the frequency range 1--20 MHz,
where the electrons in the ionosphere block our ground-based view of
the universe \cite{RadioAE2, Brown:73, Manning}.  Naturally, a space
or moon-based low-frequency radio array would not suffer from the
drawbacks of the Earth's ionosphere, as well as from the interference
noise introduced by terrestrial radio and TV stations. A number of
experiments have been proposed to measure the cosmic radiation at the
long wavelengths, which includes a network of freely flying
spacecrafts working as an antenna array \cite{Weiler:1988}, and an
antenna array installed on the moon \cite{Lunar,Lazio}. Such arrays would
not suffer from ionospheric distortions or from the radio-frequency
interference and would be able to access Galactic and extragalactic
hectometer emission, i.e. the wavelength regime corresponding to the
range at which the redshifted 21-cm signal all the way to the epoch of
recombination at $z\sim 10^3$ may be found.
 
The 21-cm signal of neutral hydrogen from the dark ages prior to star
formation \cite{Loeb:2004, 21fromDark} is expected to be extremely
weak and particularly challenging to observe. Here we revisit previous
calculations and include for the first time the effect of the UV
radiation that builds up during the cosmic recombination epoch of
hydrogen. In particular, the excess photons near the Ly$\alpha$
resonance are immediately reabsorbed by neutral hydrogen atoms,
driving the 21-cm transition out from thermal equilibrium with the
cosmic microwave background (CMB). The associated Wouthuyson-Field
(WF) effect \cite{Wouthuysen:1952, Field:1958} couples the 21-cm
emission (or absorption) to the color temperature of the Ly$\alpha$
field, boosts the magnitude of the 21-cm signal from $z>500$ and
improves its prospects for a future detection.

Our study has become possible due to recent advances in calculating
the spectral distortions of the CMB during the cosmic recombination
epoch. Two independent codes, HyRec \cite{HyRec} and CosmoRec
\cite{Chluba:2011}, were developed to accurately solve the atomic
physics together with the spectral distortions of the UV background
around Lyman-series resonances. The two codes work in a similar manner
by implementing an efficient and accurate method for solving the
multi-level atom recombination problem \cite{AliHaimoud:2010}, which
accounts for an large number of excited states. The codes include all
important processes that affect the recombination history, such as
two-photon transitions (emission, absorption, Raman scattering) from
the 2s hydrogen state and from higher levels, and frequency diffusion
in the Ly$\alpha$ line, but they ignore collisional transitions.  
The main difference between the two codes from our perspective is that
CosmoRec accounts for the effect of helium on the recombination radiation, 
while HyRec only includes the contribution from hydrogen.
The photons emitted during
helium recombination at $z\sim 2500$, contribute to the intensity of
the radiative background near the Ly$\alpha$ resonance at the lower
redshifts considered here \cite{Chluba:2009} and therefore play a role
in the WF effect. We therefore choose to use the output spectral
distortions from CosmoRec for this work.


The paper is organized as follows. In \S 2 we discuss the UV
background during and shortly after recombination and show the effect
of this background on the 21-cm signal from this epoch, in \S 3 we
remark on future observations of the hectometer signal and
conclude. Throughout our discussion we adopt the standard set of
cosmological parameters: $h = 0.6704$, $\Omega_c h^2 = 0.1204$,
$\Omega_b h^2 = 0.0220$, $\Omega_\Lambda = 0.6817$, $Y_P = 0.2477$,
$n_s = 0.9619$, $\ln(10^{10}A_s) = 3.098$, $k_0 = 0.05$ Mpc$^{-1}$
\cite{Planck:Res}. As mentioned we use CosmoRec, available
online\footnote{www.chluba.de/cosmorec},
in our calculations of the 21-cm signal.

\section{21-cm signal from $z>500$}
\subsection{The Ly$\alpha$ background}

For a fixed set of cosmological parameters, the neutral hydrogen
emission at high redshifts prior to cosmic star formation is fully
determined by atomic physics. The key processes that determine the
expected signal from this epoch are: {\it (i)} the WF effect of the
Ly$\alpha$ photons from recombination; {\it (ii)} the coupling to the
CMB, which tends to bring the hydrogen atoms to equilibrium with the
thermal bath of the CMB photons at the temperature $T_{CMB}$; and {\it
(iii)} collisions. The effect of collisions on the 21-cm signal from
dark ages is well understood \cite{Loeb:2004, 21fromDark}. The neutral
hydrogen emission from redshifts higher than $z\sim 50$, which in the
standard cosmological model marks the epoch when stars appeared for
the first time \cite{Fialkov:2011}, is completely decoupled from
complex processes associated with star formation such as gas cooling,
feedback of stellar radiation and metal enrichment, X-ray heating of
the intergalactic medium, and Ly$\alpha$ feedback, which must be taken
into consideration when making predictions for the 21-cm signal from
lower redshifts \cite{Loeb:2012}. 
At redshifts $50\lesssim z \lesssim 300$ collisions break the coupling
of the 21-cm signal to the CMB and tend to redistribute the population
of the excited, $n_1$, and ground, $n_0$, hyperfine states of the 1s
hydrogen level in accordance with the Boltzmann factor
$({g_1}/{g_0})\exp(-T_\star/T_K)$, where $T_K$ is the kinetic
temperature of the gas (which at $z>300$ is very close to the CMB
temperature), $(g_1/g_0) = 3$ being the ratio of the spin degeneracy
of the two (triplet and singlet) levels, and $T_\star = 0.068$ K the
temperature corresponding to the energy difference between the two
hyperfine levels.  This signal, considered in references \cite{Loeb:2004,
21fromDark}, is determined by the atomic physics during the early
stage of the evolution of the Universe and depends on the density of
neutral hydrogen atoms, and their collision rates with other atoms,
electrons and protons, which can be found in tabular form in the
references \cite{Furlanetto:2006}, \cite{Furlanetto:2007a} and
\cite{Furlanetto:2007b} for the hydrogen-hydrogen (H-H) ,
hydrogen-electron (H-e) and hydrogen-proton (H-p) collisions
respectively.

At yet higher redshifts, which are accurately considered in this paper
for the first time, the Ly$\alpha$ background, produced during
hydrogen recombination, plays a significant role. Understanding the
Ly$\alpha$ background is essential when making realistic predictions
for the redshifted 21-cm signal. The Ly$\alpha$ photons are absorbed
and then re-emitted by hydrogen atoms. As a result of this process an
atom may relax to a different hyperfine state than the initial one,
and the number densities of atoms in excited and ground hyperfine
states are redistributed, through the WF effect. If the WF effect
dominates the level distribution, the ratio of the number densities
of the excited and the ground hyperfine states would be $n_1/n_0 =
3\exp\left(-T_\star/T_C\right)$, with $T_C$ the color temperature of
the Ly$\alpha$ photons defined via ${T_C^{-1}} \equiv
-\frac{k_B}{h_{pl}}\frac{d\ln n_\nu}{d\nu}$ , where $k_B$ is the Boltzmann constant,
$h_{pl}$ is the Planck constant, $\nu$ is the photon frequency 
and $n_\nu = c^2J_\nu/2\nu^2$  is the photon occupation number with  $J_\nu$ 
being the intensity of the radiation field around the 
Ly$\alpha$ resonance (see Ref.  \cite{Furlanetto:2006} for more details). 
On the other hand, when the coupling $x_\alpha$ of the
Ly$\alpha$ field to the 21-cm line is weak, other processes determine
the emission signal. The coupling is defined as \be
\label{eq:xa}x_\alpha =
\frac{4P_\alpha}{27A_{10}}\frac{T_\star}{T_{CMB}},\ee where  $P_\alpha = 4\pi \sigma_0\int d\nu
J(\nu)\phi_\alpha(\nu)$ is the total scattering rate of Ly$\alpha$
photons, with  
$\phi_\alpha(\nu)$ the normalized Voigt profile, $\sigma_0 = ({\pi e^2}/{m_e c})f_\alpha$ [cm$^2$
s$^{-1}$] and $f_\alpha = 0.4875$ being the oscillator strength, and $A_{10} = 2.85\times 10^{-15}$ [s$^{-1}$] is the spontaneous decay
rate of the 21-cm line.  Since $x_\alpha$
depends on $J_\nu$, the resulting 21-cm signal is dictated by the
intensity and color temperature of the radiation background around the
Ly$\alpha$ resonance.

To find the intensity and the color temperature of the Ly$\alpha$
radiation at every instant one needs to accurately solve the
recombination problem accounting for coupled evolution of level
population of hydrogen and helium atoms and the UV radiation field,
during hydrogen and helium recombination. Here we use the results of
CosmoRec \cite{Chluba:2011}, which takes account of the important
processes during cosmic recombination (but ignores the effect of
collisions on the level population). The radiative transfer of the
Ly$\alpha$ photons is affected by: {\it (i)} emission and absorption
that are not part of scattering processes; {\it (ii)} Hubble
expansion, which shifts photons to lower frequencies; {\it (iii)}
resonant scattering of the photons off of hydrogen atoms, which leads
to energy exchange between photons and matter preserving the number of
photons; and {\it (iv)} the 2s$\rightarrow$1s two-photon decay (see
Ref. \cite{Hirata:2009} for details).  The optical depth for the
Ly$\alpha$ photons during recombination is very high; when a hydrogen
atom recombines and emits a Ly$\alpha$ photon, the photon is almost
immediately reabsorbed by another hydrogen atom.  Because of the large
difference between the 2p$\rightarrow$1s transition rate and the
Hubble rate, each photon close to the Ly$\alpha$ resonance scatters
$10^8- 10^9$ times before it can escape, finally allowing a hydrogen
atom to settle into the ground state.  This process is known to
create a bottleneck for recombination, but it also feeds the
intensity of the radiation field around the Ly$\alpha$ resonance and
thus the 21-cm signal from this epoch.  The bottleneck for
recombination is broken due to two main processes, the redshifting of
the photons out of the Ly$\alpha$ resonance and the two-photon decay
\cite{Peebles:1968}. 
As recombination proceeds, additional atoms recombine to the ground
state and the intensity of the radiation around the Ly$\alpha$
resonance declines.

\begin{figure}
\begin{center}
\includegraphics[width=3.1in]{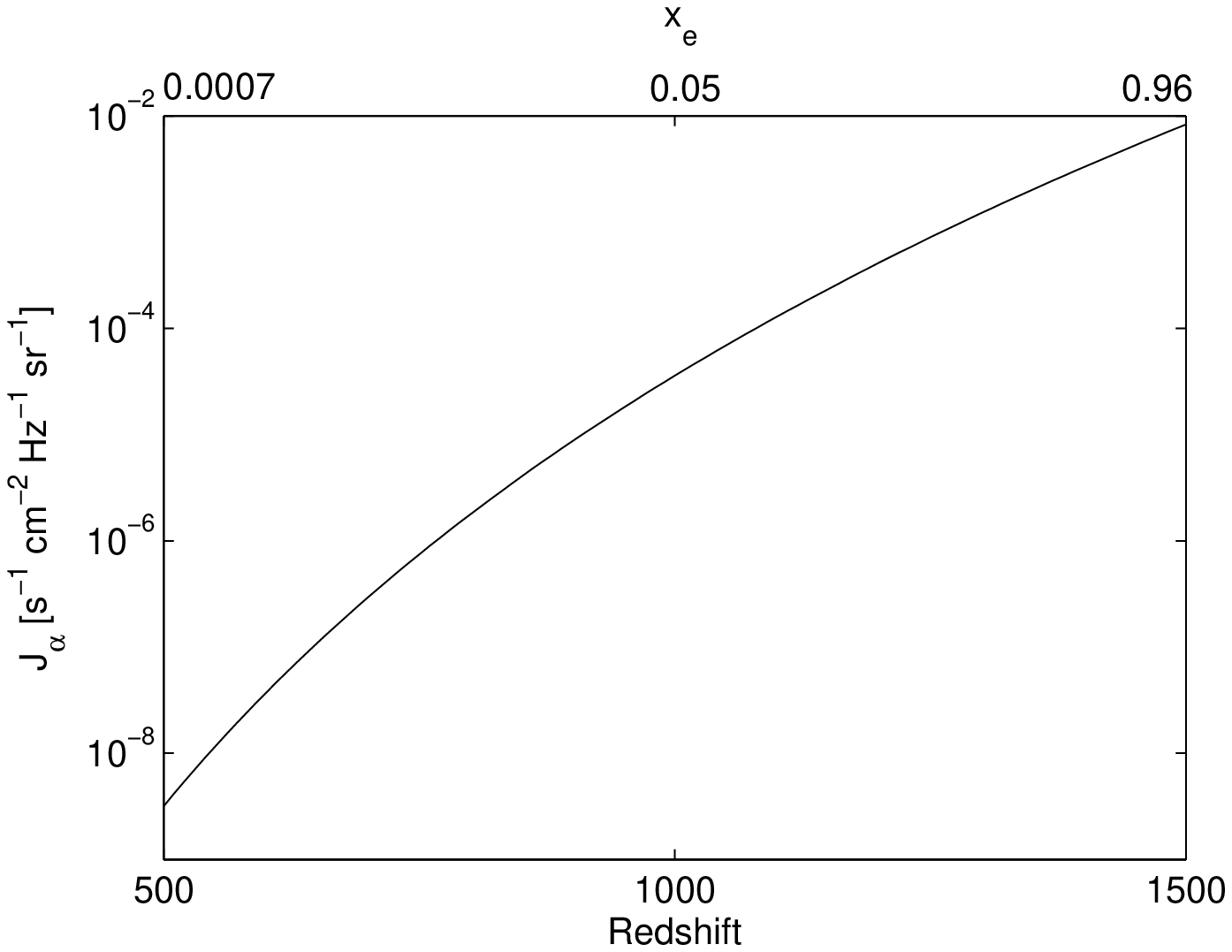}\includegraphics[width=3.1in]{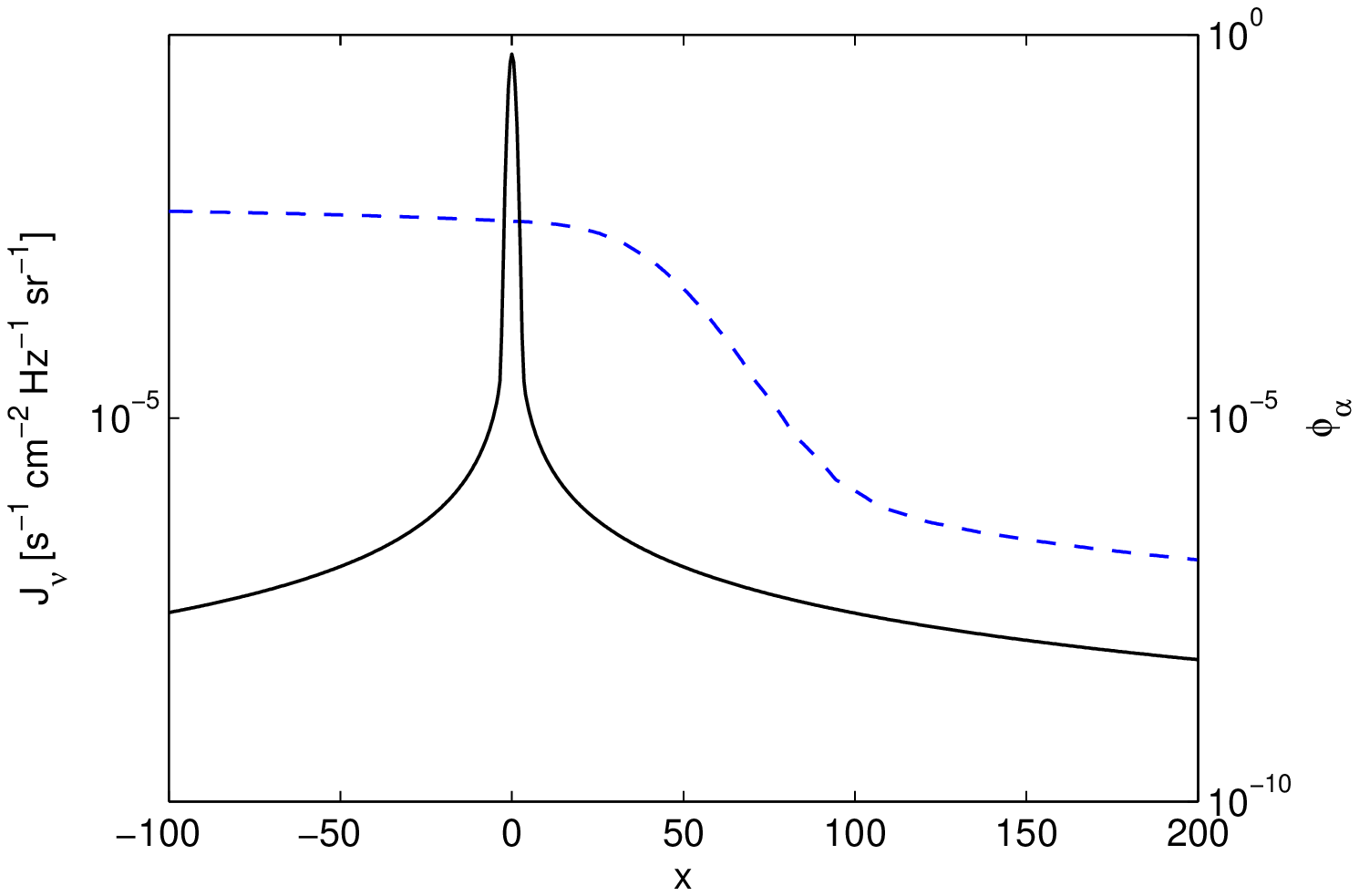}
\caption{{\it Left:} Intensity of the radiation background $J_\nu$
[s$^{-1}$cm$^{-2}$Hz$^{-1}$sr$^{-1}$] at the Ly$\alpha$ resonance,  $\nu = \nu_\alpha$, during
hydrogen recombination (compiled using the CosmoRec code) as a function of redshift (lower horizontal axis) 
and the ionization fraction $x_e$ (upper horizontal axis).  {\it Right:}
Spectral shape of $J_\nu$ [s$^{-1}$cm$^{-2}$Hz$^{-1}$sr$^{-1}$] around
the Ly$\alpha$ resonance (dashed blue line, left axis)
and the normalized Voigt profile of the Ly$\alpha$ line, $\phi_\alpha$ at $z = 1000$
(solid black line, right axis) as a function of $x \equiv
\frac{(\nu-\nu_{\alpha})}{\nu_{\alpha}}\left(\frac{2k_BT_K}{m_H
c^2}\right)^{-1/2}$, where $m_H$ is the hydrogen atom mass.}
\label{fig:Jnu}
\end{center}
\end{figure}

The left panel of Figure \ref{fig:Jnu} shows the redshift evolution of
the intensity of the radiation background $J_\nu$ (in units
s$^{-1}$cm$^{-2}$Hz$^{-1}$sr$^{-1}$) around the Ly$\alpha$ resonance.
This intensity grows monotonically with increasing lookback time
during hydrogen recombination, but is  expected to decline after the epoch of
helium recombination (HeII$\rightarrow$HeI) at $z\sim 2500$.
This Ly$\alpha$ background was our main
motivation to consider the 21-cm signal from redshifts higher than the
standard upper limit considered in the literature (e.g., $z\sim 300$
in \cite{21fromDark}). The right panel of Figure \ref{fig:Jnu} shows
the spectral shape of the intensity in the vicinity of the Ly$\alpha$
resonance and the line profile (assumed to be the normalized Voigt
profile) which are needed for the computation of $x_\alpha$ in
Equation (\ref{eq:xa}).

\subsection{Results}
The deviation of the brightness temperature of the 21-cm signal from
that of the CMB, is given by
\begin{equation}
T_b(z) = \left(1-e^{-\tau}\right)\frac{(T_S-T_{CMB})}{(1+z)},
\end{equation} 
where $\tau$ is the optical depth for the 21-cm photons, and $T_S$ is
the spin temperature defined by the ratio between the density of atoms
in the upper and lower states of the transition $(n_1/n_0) =
3\exp\left(-T_\star/T_S\right)$. To determine the spin temperature and
make predictions for the observable brightness temperature, one should
in principle evolve the number densities in time taking account of the
Hubble expansion rate, the radiative transition rates due to the
interation with the CMB, the collisional spin excitation and
de-excitation rates, and spin excitation and de-excitation rates from
Ly$\alpha$ absorption \cite{Loeb:2004}. However, at the redshifts of
interest, the Hubble rate is sub-dominant compared to the other rates.
As a result, we can safely assume that the system reaches a radiative
equilibrium, in which case the spin temperature is given by \cite{Furlanetto:2006}
\be\label{eq:TS}\frac{1}{T_s} = \frac{T_{CMB}^{-1}+x_C
T_K^{-1}+x_\alpha T_C^{-1}}{1+x_C+x_\alpha},\ee where $x_C$ is the
collisional coupling.

\begin{figure}
\begin{center}
\includegraphics[width=3.1in]{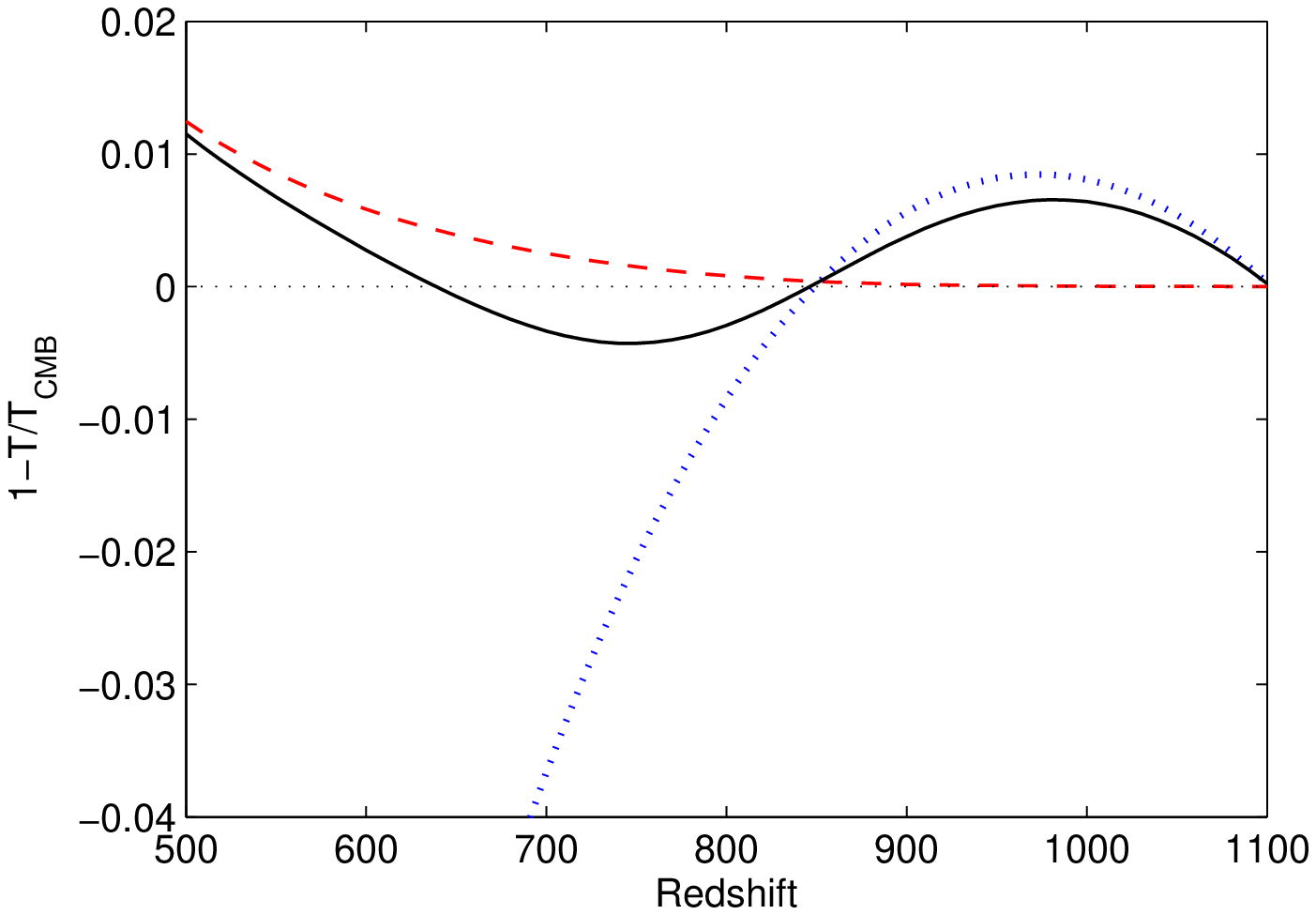}\includegraphics[width=3.1in]{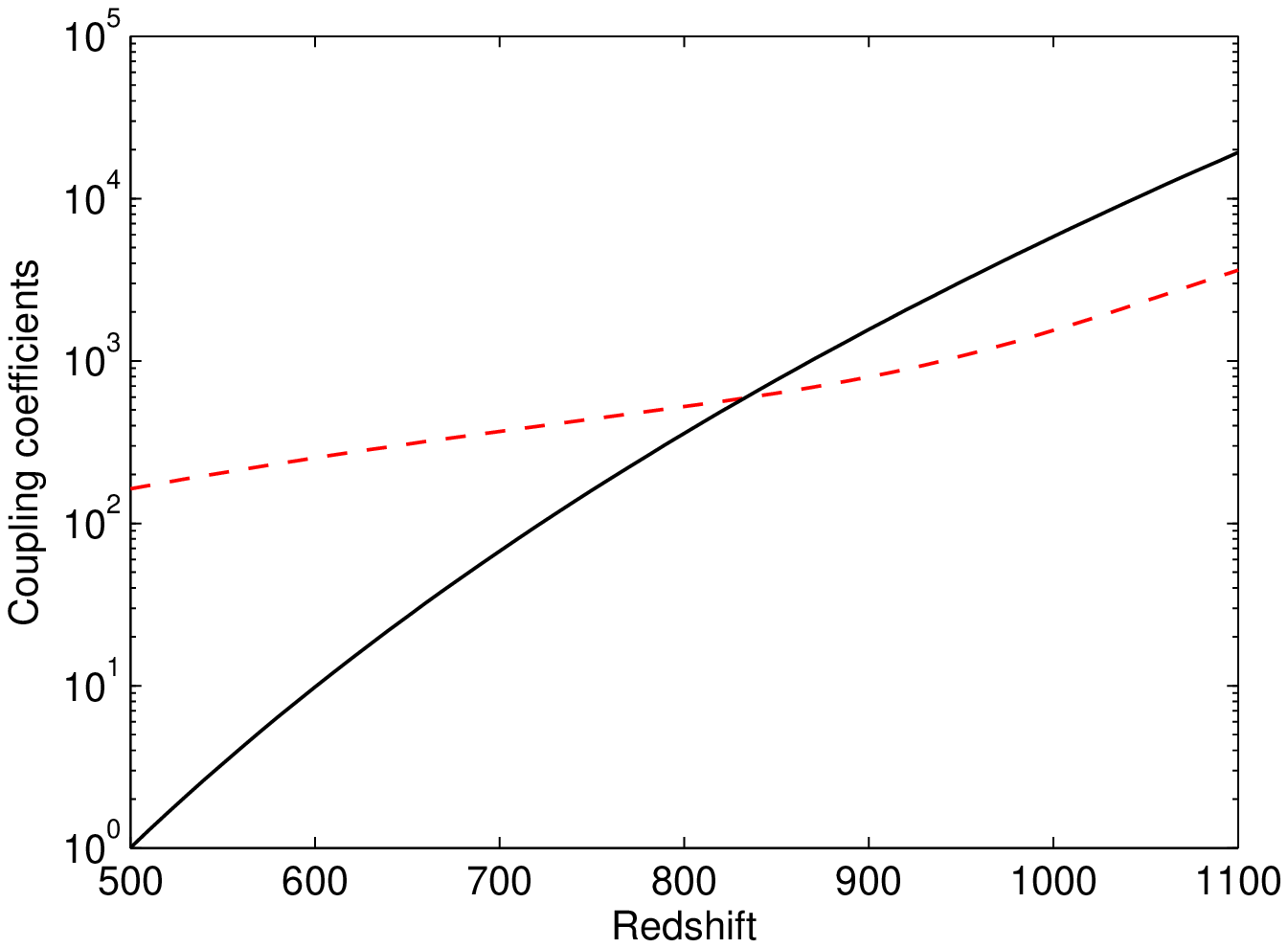}
\caption{{\it Left:} Reshift evolution for the fractional contrast factors of various
temperatures $T_i$ relative to the CMB temperature,
$(1-{T_i}/{T_{CMB}})$. We show the kinetic gas temperature $T_i=T_K$
(red dashed), spin temperature $T_i=T_S$ (black solid) and color
temperature $T_i=T_C$ (blue dotted). {\it Right:} Coupling
coefficients as a function of redshift. We show the Wouthuysen-Field
coupling coefficient (black solid curve) and the collisional coupling
including H-H, H-p and H-e collisions (red dashed curve). The
Wouthuysen-Field effect becomes dominant at $z\gtrsim 850$.}
\label{fig:TK} 
\end{center}
\end{figure}

The dominant contribution to the brightness temperature of the 21-cm
signal in the redshift range $40\lesssim z \lesssim 850$ originates
from the collisional coupling.  At higher redshifts this contribution
alone vanishes due to the thermal coupling between the gas temperature
and the CMB.  However, when the effect of the early Ly$\alpha$ photons
is added, at high redshifts the spin temperature couples to the the color temperature of
Ly$\alpha$.  On the left panel of Figure \ref{fig:TK} we compare
between the factor $(1- {T_{S}}/{T_{CMB}})$, which contributes to the
brightness temperature, and its limits at low redshifts
$(1-{T_{K}}/{T_{CMB}})$ (the case of saturated collisional coupling)
and at high redshifts $(1-{T_{C}}/{T_{CMB}})$ (the case of saturated
WF coupling). The rapid evolution of the color temperature, shown on the Figure, reflects
the fact that the slope of the radiation field around the Ly$\alpha$
resonance depends strongly on redshift and is determined by the
aspects of atomic physics during hydrogen and helium recombination in
addition to the cosmic expansion.  The right panel of Figure
\ref{fig:TK} shows the relative importance of $x_\alpha$ and $x_C$ as
a function of redshift. As is evident, collisional coupling dominates
at redshifts below $z\sim 850$ while $x_\alpha$ dominates at higher
redshifts. Thus, it is essential to account for the early WF effect
when making predictions for the 21-cm signal from the epoch of
recombination.

\begin{figure}
\begin{center}
\includegraphics[width=3.1in]{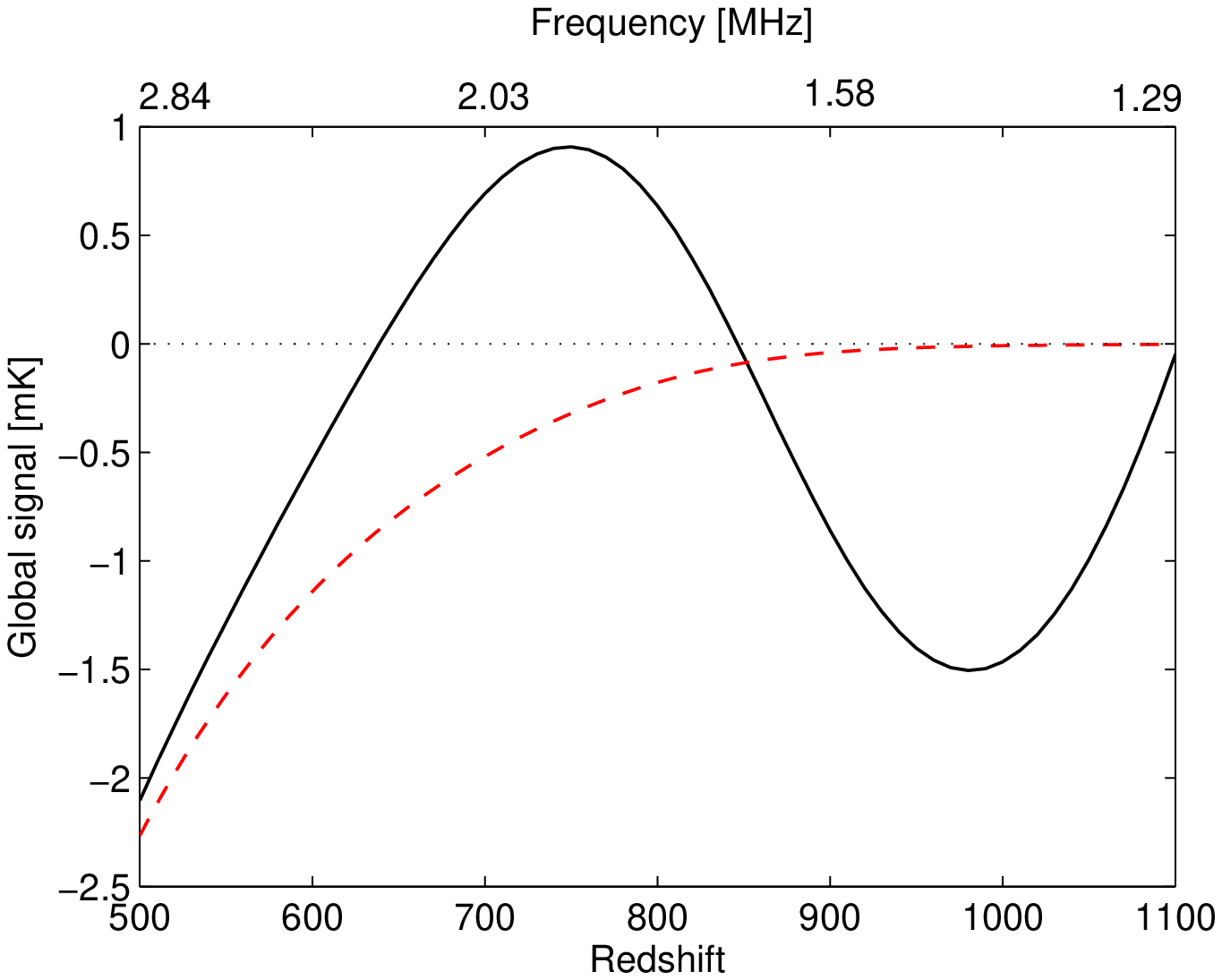}\includegraphics[width=3.1in]{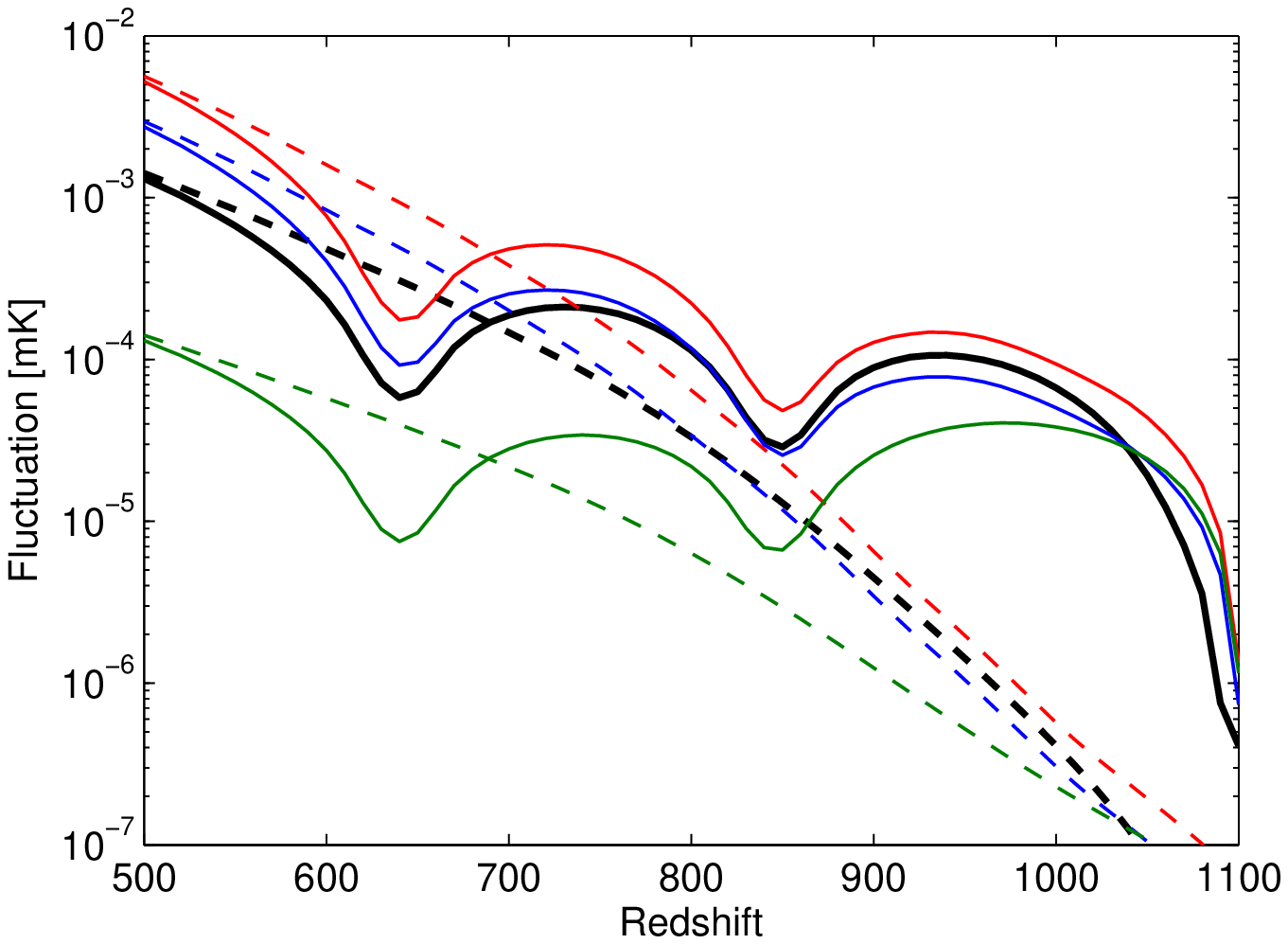}
\caption{{\it Left:} The amplitude of 21-cm brightness temperature in
mK when accounting only for collisions (dashed, red) and when adding
the effect of the Ly$\alpha$ photons (solid black). The Ly$\alpha$
contribution is noticeable at $z>500$. {\it Right:} The fluctuations
in the 21-cm background in mK (defined via Eq. \ref{eq:dTb}) versus redshift including the WF
coupling $x_\alpha$ (solid curves) and ignoring it (dashed curves) for comoving
wavenumbers $k = 5$ Mpc$^{-1}$ (red), $k = 0.5$ Mpc$^{-1}$ (blue), the
Baryonic Acoustic Oscillations (BAO) scale $k = 0.05$ Mpc$^{-1}$
(boldface, black), and $k = 5$ Mpc$^{-1}$ (green).}
\label{fig:T21} 
\end{center}
\end{figure}

The global spectrum of neutral hydrogen, representing the average
21-cm brightness across the sky, is shown on the left panel of Figure
\ref{fig:T21}. Its shape reflects the combined effects of collisional
coupling and Ly$\alpha$ photons in addition to the radiative coupling
to the CMB. The presence of the early Ly$\alpha$ photons leads to an
enhancement in the 21-cm signal from high redshifts ($z>500$) at low
observed frequencies ($\nu< 2.84~{\rm MHz}$). Since the color
temperature, and hence the spin temperature at these redshifts,
changes its behavior relatively to $T_{CMB}$ with redshift (as shown
in Figure \ref{fig:TK}), the signal is expected to be seen in emission
at $640\lesssim z \lesssim 850$ and in absorption at $850\lesssim z
\lesssim 1100$.  The amplitude of the emission signal peaks at
$z\sim750$ where it reaches a value of $T_b = 0.9$ mK, instead of
$-0.35$ mK in the collisions-only case previously considered in the
literature. The minimum of the absorption signal during the
recombination era is reached at $z\sim 980$ with a value of $T_b =
-1.5$ mK instead of $-0.01$ mK in the collisions-only case. Note that
since the intensity of the radiation field around Ly$\alpha$
resonance is well defined by atomic physics (calculated here through
the CosmoRec code), one should be able to predict the signal precisely
for a fixed set of cosmological parameters.

Despite the growing intensity of Ly$\alpha$ photons towards higher
redshifts (as shown on Figure \ref{fig:Jnu}), the 21-cm signal from
$z>1060$ diminishes, since the optical depth for free-free absorption
of 21-cm photons exceeds unity. This defines a fundamental cutoff in
redshift space beyond which the 21-cm signal cannot be observed.  Note
that the Thompson optical depth of redshifted 21-cm signal is smaller
than $10\%$ at all redshifts of interest, implying that
observations of the 21-cm signal at redshifts below the cutoff (and
above the redshift of reionization) are in principle possible.

Next, we consider the effect of the early Ly$\alpha$ photons on the
power spectrum of the 21-cm signal. The WF effect is essential in
maintaining the 21-cm signal at high redshifts, as shown by the right
panel of Figure \ref{fig:T21}. During and after recombination but
before formation of first stars, the main source of fluctuations is
perturbations in the neutral hydrogen density, $\delta n_{HI}/n_{HI}$
\cite{Loeb:2004}. Since the ionization fraction during recombination
is expected to be nearly uniform, this quantity follows the baryon
density perturbations $\delta_b\equiv ({\delta \rho_b}/{\rho_b})$, which
we here find using the online version of the CAMB code
\cite{Lewis}. Ignoring other sources of fluctuations, the wavenumber
dependence of the 21-cm power spectrum is that of the baryonic density
field, with the early WF coupling affecting its redshift-dependent
amplitude. In this regime, perturbations in $T_b$ trace perturbations
in hydrogen density $\delta T_b = \frac{d T_b}{d n_{HI}}\delta
n_{HI}$, and we can analytically derive the fluctuations in the 21-cm
field to be, \be \label{eq:dTb}\delta T_b
= T_b\sqrt{1.87}\left[\frac{2+x_\alpha+x_c}{1+x_\alpha+x_c}\right]\delta_b,\ee where we accounted for the dependence of $\tau$,
$x_C$ and $x_\alpha$ on $n_H$. The factor $\sqrt{1.87}$ 
accounts for the spherically-averaged contribution of peculiar velocities
to the fluctuation amplitude
of the brightness temperature within  linear theory \cite{Barkana:2005b}. (In the case when the WF coupling is ignored, this equation becomes $ \delta T_b
= T_b\sqrt{1.87}\left[\frac{2+x_c}{1+x_c}\right]\delta_b$.)
The right frame of Figure \ref{fig:T21} shows the redshift dependence
of the 21-cm fluctuations at particular comoving wavenumbers for the
two cases: with and without the early Ly$\alpha$ background assuming a
standard set of cosmological parameters \cite{Planck:Res}. In general,
fluctuations from redshifts $z>500$ are very small compared to the
gravitationally-amplified fluctuations at lower redshifts (for
example, the maximal fluctuation from collisions reach the $\sim 0.6$
mK level at the BAO wavenumber scale at $z\sim50$), mainly due to the
smallness of the density perturbations in baryons at $z\sim 10^3$
whose growth was suppressed by the CMB before recombination. However
Ly$\alpha$ photons prevent the 21-cm signal at high redshifts from
decaying and keep it at a nearly constant amplitude of
$\mathcal{O}(10^{-5}-10^{-4})$ mK, while the previously calculated
(collisions-only) amplitude decays exponentially fast with increasing
redshift. The ratio of the two signals is scale independent and peaks
at $z\sim 1060$ (the redshift after which the 21-cm signal drops due
to free-free absorption) where it reaches a value of 230, with a
secondary peak in the ratio obtained at a lower redshift $z \sim
790$ with a value of $3.5$.

\section{Conclusions}

We performed the first calculation of the effect of the Ly$\alpha$
radiation background during the epoch of recombination on the 21-cm
signal from cosmic hydrogen. We found that the Wouthuysen-Field
coupling of the 21-cm signal to the color temperature of the
Ly$\alpha$ radiation elevates the global 21-cm signal at observed
frequencies $\nu <2.84$ MHz (or wavelengths $\lambda >100$ meters), to
a milli-Kelvin level in the redshift range $z\sim 500-1100$. The power
spectrum of 21-cm brightness fluctuations obtains an amplitude of
order $(10^{-5}$--$10^{-4})$ mK, well above previous predictions which
ignored the Ly$\alpha$ background.  At redshifts higher than $z\sim
1060$ ($\nu <1.34$ [MHz] or $\lambda> 222$ meters) the cosmic plasma
becomes opaque to the 21-cm signal due to free-free absorption.

Even though the Ly$\alpha$ coupling greatly increases the expected
21-cm signal from the hydrogen recombination epoch, detection of the
hectometer waves, which must be done from space with a lunar or a
free-flying antenna array, still appears very futuristic. However, ongoing
interest in the ultra-long wavelength regime which was largely
unexplored so far, as well as technological advances in
instrumentation, might eventually lead to the construction of related
observatories and possibly even to the detection of the global 21-cm
signal in the 1.5--3 MHz band. To access the cosmic signal, a thorough
study of the Galactic emission at the very low frequencies must be
conducted.  The first observations in the frequency range of 0.4--6.5
MHz were done in the 1970s by the Radio Astronomy Explorer satellite
\cite{RadioAE2}. The spectrum measured by this satellite was dominated
by the Galactic absorption and emission and reached the maximal
brightness temperature of $\sim 4\times10^{9}$ mK at 3 MHz, which is
nine orders of magnitude stronger than the cosmic signal predicted
here.  This result was later confirmed by measurements of the spectrum
at 0.2--13.8 MHz by the WAVES instrument on the WIND spacecraft
\cite{Manning}. Hopefully, the technology developed at present for the
reconstruction of the 21-cm signal from lower redshifts, where the
Galactic emission is ``only'' $\sim 10^5$ stronger than the cosmic
signal, will pave the way to the detection of the 21-cm from
recombination. Proposals for radio astronomy on the moon \cite{Lunar}
suggest observations in the frequency range 1--10 MHz as one of their
primary goals. As a first step, it is being suggested to install a
single lunar antenna operating in the frequency range of 1 kHz--100
MHz, which could serve as a pathfinder for a future lunar array to
detect the 21-cm signal from high redshifts.  One of the goals of such
an array would be to map the signal at frequencies inaccessible from
the ground which are completely reflected by the ionosphere of the
Earth, i.e.  1--10 MHz, matching the range of frequencies that define
the 21-cm signal from recombination. The other band of interest is
10--100 MHz, where the incoming phase of the radiation is distorted by
the ionosphere. Currently, there exists an unfunded ESA proposal to launch 
a Lunar Lander with such a path-finder  antenna on board. Another proposed project involves the construction of a radio observatory on the lunar surface \cite{Lazio}. Although its primary goal concerns imaging of the radio emission generated by solar bursts at frequencies below 10 MHz and studying the lunar ionosphere, the proposed observatory is also designed to serve as a path-finder for a larger array on the moon focused on cosmology.

If observed, the 21-cm signal from $z>500$ will provide a unique,
independent and rich probe of the early Universe. For a fixed set of
cosmological parameters and no new physics, the expected signal
depends solely on atomic physics and is not ``contaminated'' by the
products of star formation. Therefore, it offers a unique
three-dimensional probe (as opposed to the CMB which probes only the
two-dimensional surface of last scatter) of cosmological parameters
and primordial density perturbations, allowing refined tests of the
$\Lambda$CDM model, the properties of the dark matter and dark energy,
as well as theories of modified gravity and inflationary
cosmology. Since the 21-cm signal does not suffer from Silk damping on
small scales, it allows to set new constrains on the spectrum of
isocurvature perturbations on extremely small scales (down to the
Jeans length \cite{Loeb:2004}) where a blue tilt \cite{Isocurve} would
be manifested.

Our study was made possible by recent advances in the theory of
recombination, encapsulated by the codes CosmoRec and HyRec which
calculate the radiative transfer of the Ly$\alpha$ line.  We found
order unity differences between the predictions for the 21-cm signal
from the two available codes, and hence conclude that a better
treatment of the Ly$\alpha$ background is needed in order to reach the
level of precision characterizing the current standard for predictions
of CMB anisotropies.

\acknowledgments 

We thank Y. Ali-Haimoud and J. Chluba for useful discussions about the
HyRec and CosmoRec codes. This work was sponsored by the Raymond and Beverly Sackler
Tel-Aviv University Harvard/ITC Astronomy Program, additional support was obtained from Israel Science Foundation grant 823/09,  the LabEx
ENS-ICFP: ANR-10-LABX-0010/ANR-10-IDEX-0001-02 PSL and NSF grant
AST-1312034.

\end{document}